\newcommand{\NPA}[3]{Nucl.\ Phys.\ {\bf A#1},\ #2 (#3)}
\newcommand{\NPB}[3]{Nucl.\ Phys.\ {\bf B#1},\ #2 (#3)}

\newcommand{\PLB}[3]{Phys.\ Lett.\ B\ {\bf #1},\ #2 (#3)}

\newcommand{\PRL}[3]{Phys.\ Rev.\ Lett.\ {\bf #1},\ #2 (#3)}
\newcommand{\PRA}[3]{Phys.\ Rev.\ A\ {\bf #1},\ #2 (#3)}
\newcommand{\PRB}[3]{Phys.\ Rev.\ B\ {\bf #1},\ #2 (#3)}

\newcommand{\PRD}[3]{Phys.\ Rev.\ D\ {\bf #1},\ #2 (#3)}
\newcommand{\JPG}[3]{J.\ Phys.\ G\ {\bf #1},\ #2 (#3)}

\renewcommand\k{\kappa}

\newcommand{\diracslash}[1]{#1\llap{/\kern2pt}}

\newcommand{\be}{\begin{equation}}
\newcommand{\ee}{\end{equation}}
\newcommand{\bea}{\begin{eqnarray}}
\newcommand{\eea}{\end{eqnarray}}
\newcommand{\ba}[1]{\begin{array}{#1}}
\newcommand{\ea}{\end{array}}

\documentclass[epj]{svjour}
\usepackage{graphics}
\begin{document}
\title{LOFF and breached pairing with cold atoms }
\author{Amruta Mishra\inst{1}\and
Hiranmaya Mishra\inst{2}\inst{3}}

\institute{Department of Physics, Indian Institute of Technology, New 
Delhi-110016,India
\email amruta@physics.iitd.ac.in
\and
{Theory Division, Physical Research Laboratory,
Navrangpura, Ahmedabad 380 009, India}
\and
{School of Physical Sciences, Jawaharlal Nehru University, New Delhi-110 067, India}}


\def\be{\begin{equation}}
\def\ee{\end{equation}}
\def\bearr{\begin{eqnarray}}
\def\eearr{\end{eqnarray}}
\def\zbf#1{{\bf {#1}}}
\def\bfm#1{\mbox{\boldmath $#1$}}
\def\hf{\frac{1}{2}}
\def\kp{\zbf k+\frac{\zbf q}{2}}
\def\km{-\zbf k+\frac{\zbf q}{2}}
\def\hwo{\hat\omega_1}
\def\hwt{\hat\omega_2}
\abstract
{We investigate here the Cooper pairing of fermionic atoms with mismatched
fermi surfaces using a variational construct for the ground state. We
determine the state for different values of the mismatch of chemical potential
for weak as well as strong coupling regimes including the BCS BEC cross 
over region. We consider Cooper pairing with both zero and finite net 
momentum. 
Within the variational approximation for the ground state and comparing the
thermodynamic potentials, we show that (i) the LOFF phase is stable in the 
weak coupling regime, (ii) the LOFF window is maximum on the BEC side near 
the Feshbach resonance and (iii) the existence of stable gapless states 
with a single fermi surface for negative average chemical 
potential on the BEC side of the Feshbach resonance. 
}
\maketitle

 \section{Introduction}
\label{intro}

Fermionic superfluidity driven by s-wave short range interaction 
with mismatched fermi surfaces 
has attracted attention recently both theoretically \cite{gaplessth}
 as well as experimentally \cite{gaplessexp}.
The experimental techniques developed during the last few years, at present,
 enable one to
prepare atomic systems of different compositions, densities, coupling strengths 
as well as the sign of the coupling in a controlled manner with trapped
 cold fermionic atoms using techniques of Feshbach resonance \cite{expbcsbec}. 
Because of the great flexibility of such systems, the knowledge acquired in 
such studies shows promise to shed light on topics outside the realm of atomic
physics including dense quark matter in the interior of neutron stars
\cite{quarkmatter}, high $T_c$ superconductors \cite{htc} as well as for the
physics of dilute neutron gas \cite{gas}.

In degenerate fermi systems, it is well known that an attractive interaction 
destabilizes the fermi surface. Such an instability is cured by
the  BCS mechanism characterized by pairing between fermions with
opposite spins and opposite momenta near the fermi surface leading to a gap
in the spectrum. In this system the number densities and the chemical
potentials of the two condensing fermions are the same. On the other hand,
there are situations where the fermi momenta of the two species that condense
need not be the same. Such mismatch in fermi surfaces can be realized
 in different physical systems e.g. superconductor in an
external magnetic field or a strong spin exchange field \cite{loff,takada}, 
mixture of two species of fermionic atoms with different densities and/or masses
\cite{gaplessexp} or charge neutral quark matter that could be there in the
interior of neutron stars. The ground state of such systems shows possibilities
of different phases which include the gapless superfluid phase 
\cite{wilczek,rapid,igor} with the order parameter as non zero but the
excitation energy becoming zero, the LOFF phase where the order parameter 
acquires a spatial variation \cite{loff,loffitaly,loffaust,zhuang}.
 The LOFF phase 
might also manifest in a crystalline structure \cite{crystal}. 

When the mismatch in chemical potential is small as compared to the superfluid gap,
the superfluid state does not support the gapless single particle excitations.
When the chemical potential difference is large as compared to the superfluid gap
there could be possibility of breached pairing superfluidity \cite{wilczek,rapid},
where there are two fermi surfaces with the single particle excitation vanishing at
two values of the momentum. When the mass difference between the two species
is large it could also lead to interior gap superfluidity where the quasi
particle excitation vanishes at a single momentum \cite{silotri}. 
The stability of such
configurations has been studied
demanding positivity of the Meissner masses \cite{zhuang,rischkeshov} as well
as the number susceptibility\cite{andreas}. Such an analysis has also been
done for relativistic fermions with four fermi interactions \cite{rischkeshov}.
In these analyses the ground state was not considered explicitly. Instead, 
the chemical potential and the gap functions were treated as parameters, with 
the whole parameter space being analyzed systematically for stability criteria which
were related to different possible quasi particle dispersions  
\cite{rischkeshov,andreas}.

          Our approach to such problems has been variational with an explicit
construct for the ground state \cite{rapid,hmspm,hmparikh,amhma,amhmb}.
The minimization of the thermodynamic potential with respect to the functions used to
describe the ground state decides which  phase would be favored at what density 
and/or coupling. This has been applied to quark matter with charge neutrality condition
\cite{amhma,amhmb,amhmc} as well as to the systems of cold fermionic atoms \cite{rapid}.
In the present paper we would like to extend the method to have the ansatz
state general enough to include the LOFF state with a fixed momentum for the
condensate. Thus our approach will be complementary to that of Ref.\cite{andreas}
in the sense that we shall use explicit state and solve the gap equation to determine
gap function. Comparison of thermodynamic potential for
different phases after the gap equation is solved will decide which phase will be
favored at what coupling, density and the mismatch in fermi momentum. We also take
the analysis further as compared to that in Ref.\cite{andreas} to include the
LOFF state, with a single plane wave.

For the case of vanishing mismatch in chemical potentials of the
pairing species, a functional integral formulation with
a saddle point approximation was considered to describe the crossover from
BCS to Bose superfluidity \cite{randeria,marini}.
 A systematic analysis of the effect
of gaussian fluctuations around the saddle point solution was also performed.
It was shown that while the corrections from the fluctuations are extremely 
important for temperatures close to the critical temperature, particularly for 
strong coupling, the same is very small for all couplings for temperatures 
small compared to the critical temperature. The ability of this saddle point
solution which at first sight might have been expected to work only for small 
couplings, to reproduce the strong coupling Bose limit is reassuring. This
gives some confidence in its validity for intermediate coupling results 
where no obvious small expansion parameter is known 
\cite{randeria}. As we shall show in the following, the results arising 
from our ansatz for the symmetric case of equal chemical potentials 
corresponds to this saddle point solution.

We organize the paper as follows. In section \ref{ansatz}, 
we briefly introduce the
model and the discuss the ansatz for the ground  state.
In subsection \ref{evaluation}, we compute the expectation value of the
Hamiltonian  with respect to the ansatz state to calculate the
thermodynamic potential. Minimizing the thermodynamic potential, we
determine the ansatz functions in the ground state and the resulting 
superfluid gap equations. In section \ref{results},  we first discuss
various possible phases and the corresponding
results of the present investigation. Finally, we summarize 
and conclude in section \ref{summary}.
\section{ The ansatz for the ground state}
\label{ansatz}
 To discuss the superfluidity for fermionic atoms, we 
consider a system of two species of non-relativistic fermions
with chemical potentials $\mu_1$ and $\mu_2$ and having equal masses.
 These two species can e.g. be the two hyperfine states of $^{40}Ca$.
Further we shall assume a point like interaction between the two species.
The Hamiltonian density is given as
\be
{\cal H}(\zbf r)=
\psi_a(\zbf r)^\dagger(-\frac{\zbf\nabla^2}{2m}-\mu_a)\psi_a(\zbf r)+
g\psi^\dagger_1(\zbf r)\psi_1(\zbf r)\psi_2(\zbf r)^\dagger \psi_2(\zbf r)
\label{ham}
\ee
where, $\psi^a(\zbf r)$ is the annihilation operator of the fermion species 
`$a$',
and, $g$ is the strength of interaction between the two species of the fermionic atoms.
Throughout the paper we shall work in the units of $\hbar =c=k_B=1$.

While considering Cooper pairing between fermions with different fermi 
momenta, it was realized by Larkin - Ovchinnikov and Fulde- Ferrel (LOFF), 
that it could be favorable to have Cooper pairing with nonzero total momentum 
unlike the usual BCS pairing of fermions with equal and opposite momentum. 
We shall examine here whether an ansatz with a condensate of Cooper pair 
of fermions having momenta $\zbf k+\zbf q/2$ and $-\zbf k+\zbf q/2$, 
thus with a nonzero net momentum $\zbf q$
of each pair is favored over either BCS condensate or the normal state without
condensates. Here, our notation is such that $\zbf k$ specifies a particular
Cooper pair while $\zbf q$ is a fixed vector, the same for all pairs which
characterizes a given LOFF state. Here the magnitude of the momentum $|\zbf q|$
shall be determined by minimizing the thermodynamic potential while the 
direction is chosen spontaneously.

With this in mind, to consider the ground state for this system of
fermionic atoms with  mismatched fermi surfaces ($\mu_1\neq \mu_2$), 
we take the following ansatz for the ground state as 
\be
|\Omega\rangle={\cal U}_q|vac\rangle
\ee
where, $|vac\rangle$ is annihilated by $\psi_r(\zbf r)$ and the unitary
 operator ${\cal U}_q $ is given as
\begin{equation}
{\cal U}_q=\exp\left (
\int \psi^a(\zbf k+\frac{\zbf q}{2})^\dagger
\psi^b(-\zbf k+\frac{\zbf q}{2})^\dagger
\epsilon_{ab} f(\zbf k)d\zbf k
 -{\rm {h.c.}}\right ).
\label{u1}
\end{equation}
In the above, $\zbf k$ identifies a fermionic pair and 
each pair in the condensate has the same 
net momentum $\zbf q$. The function $f (\zbf k)$ is a variational function
to be determined through the minimization of the thermodynamic potential.
In the limit of zero net momentum ($\zbf q =\zbf 0$) this ansatz reduces to
the BCS ansatz considered earlier \cite{rapid,amhmb,amhmc}. Such an ansatz 
breaks translational and rotational invariance. In position space,
as we shall show that it describes a condensate which varies as a plane
wave  with wave vector $\zbf q$.
To include the effect of temperature and density we write down the state 
at finite temperature and density,
$|\Omega(\beta,\mu)\rangle$,  taking
a thermal Bogoliubov transformation over the state $|\Omega\rangle$ 
using thermofield dynamics (TFD) method as \cite{tfd,amph4}
\begin{equation} 
|\Omega(\beta,\mu)\rangle={\cal U}_{\beta,\mu}|\Omega\rangle={\cal U}_{\beta,\mu}
{\cal U}_q |0\rangle,
\label{ubt}
\end{equation} 
where ${\cal U}_{\beta,\mu}$ is given as
\begin{equation}
{\cal U}_{\beta,\mu}=e^{{\cal B}^{\dagger}(\beta,\mu)-{\cal B}(\beta,\mu)},
\label{ubm}
\end{equation}
with 
\begin{equation}
{\cal B}^\dagger(\beta,\mu)=\int \Big [
{\psi^\prime}^a (\zbf k)^\dagger \theta_-^a(\zbf k, \beta,\mu)
{{\underline \psi}^{\prime}}^a (-\zbf k)^\dagger 
\Big ] d\zbf k.
\label{bth}
\end{equation}
In Eq.(\ref{bth}) the ansatz functions $\theta_{-}^a(\zbf k,\beta,\mu)$,
as we shall see later, are
related to the distribution function for the $a$-th species 
of fermions,
and, the underlined operators are the operators in the extended Hilbert space 
associated with thermal doubling in TFD method. 
 All the functions in the ansatz in Eq.(\ref{ubt}), the condensate 
function $f(\zbf k)$, the distribution functions $\theta^{i}(\zbf k,\beta,\mu)$ 
are obtained by the minimization of the thermodynamic potential.
We shall carry out this minimization in the next subsection.

\subsection{Evaluation of thermodynamic potential and the gap equation }
\label{evaluation}
Having presented the trial state Eq.(\ref{ubt}), we now proceed to
minimize the expectation value of the thermodynamic potential
with respect to the variational functions $f(\zbf k)$,
$\theta^a(\zbf k,\beta)$ in the ansatz.
Using the  fact that, the variational ansatz state in Eq.(\ref{ubt}) 
arises from successive Bogoliubov transformations, 
one can calculate the expectation values
of various operators. In particular, for the 
condensate we have
\bearr
&&\langle\Omega(\beta,\mu)|\psi^a(\zbf r)\psi^b(\zbf r)|\Omega(\beta,\mu)
\rangle\nonumber\\
&&=\frac{\epsilon^{ab}\exp(i\zbf q \cdot\zbf r )}{2 (2\pi)^3}
\int \sin 2f \left(\zbf k-\frac{\zbf q}{2}\right)\nonumber\\
&&\times
\left[\sin^2\theta^b(-\zbf k+\zbf q)-\cos^2\theta^a(\zbf k)\right] d\zbf k
\nonumber\\
&&\equiv
\epsilon^{ab}\exp(i\zbf q\cdot r)I_q,
\label{iq}
\eearr
which for nonzero $I_q$ describes a plane wave with a wave vector $\zbf q$.
Once one has demonstrated the instability to formation of a single plane wave, it is
natural to expect that the state which actually develops has a crystalline
structure. LOFF, in fact, argued the favored configuration to be a crystalline
condensate which varies in space like a one dimensional standing wave 
$\cos (\zbf q\cdot \zbf r)$ with the condensate vanishing at the nodal planes.
Which crystal structure will be free energetically most favored is still to be
resolved.
The present variational ansatz Eq.(\ref{ubt}) is a first step in this direction
to decide which crystalline structure  will finally be chosen through 
minimization of the thermodynamic potential.

Similarly the number operator expectation value for the atoms of species 
`a' is given as
\bearr
\rho^a&=&\langle\Omega(\beta,\mu)|\psi^a(\zbf r)^\dagger\psi^a(\zbf r)
|\Omega(\beta,\mu)\rangle\nonumber\\
&=& \frac{1}{(2\pi)^3}\int\big[
\cos^2f(\zbf k -\frac{\zbf q}{2})\sin^2\theta^a(\zbf k)
\nonumber\\ &+&|\epsilon^{ab}|
\sin^2f(\zbf k -\frac{\zbf q}{2})\cos^2\theta^b(-\zbf k+\zbf q)
\big] d\zbf k,
\label{rhoa}
\eearr
which is independent of the space coordinate $\zbf r$.

Using Eq.s (\ref{iq}) and (\ref{rhoa}) it is straightforward to
calculate the expectation value of the Hamiltonian given in Eq.(\ref{ham}). We
obtain

\bearr
&&\epsilon
\equiv
\langle
\Omega(\beta,\mu)|{\cal H}|\Omega(\beta,\mu)\nonumber\\
&& = \frac{1}{(2\pi)^3}\int d\zbf k
\bigg[
\left(\epsilon(\zbf k+\frac{q}{2})-\mu_1\right)\nonumber\\&&
\bigg(\cos^2 f(\zbf k)\sin^2\theta_1(\zbf k +\frac{\zbf q}{2})
+
\sin^2 f(\zbf k)\cos^2 \theta_2(-\zbf k+\frac{\zbf q}{2})\bigg)
\nonumber\\
&&+
\left(\epsilon(-\zbf k+\frac{q}{2})-\mu_2\right)\nonumber\\&&
\bigg(\cos^2 f(\zbf k)\sin^2\theta_2(-\zbf k +\frac{\zbf q}{2})
\sin^2 f(\zbf k)\cos^2 \theta_1(\zbf k+\frac{\zbf q}{2})\bigg)\bigg]
\nonumber\\
&&+g I_q^2+g\rho_1\rho_2
\label{eps}
\eearr
where, the first term arises from the expectation value of the
kinetic part of the Hamiltonian and the last two terms arise from the 
four point contact interaction term. Here,
$\epsilon (\zbf k)=\zbf k^2/(2m)$, and,
the quantities $I_q$, $\rho^a$ ($a=1,2$) are defined in
Eq. (\ref{iq}) and Eq. (\ref{rhoa}) respectively. The thermodynamic
 potential is
given as
\be
\Omega=\epsilon - \mu_a \rho^a-\frac{1}{\beta}s
\label{Omega}
\ee
where, we have denoted $s$ as the total entropy density of the two 
species given as \cite{tfd,caldas}

\be
s  =  -\frac{1}{(2\pi)^3}
\sum_a\int d \zbf k
\left( \sin^2\theta^a \ln (\sin^2\theta^a) +\cos^2\theta^a
\ln (\cos^2\theta^a) \right).
\label{ent}
\ee

We now apply the variational method to determine the
condensate function $f(\zbf k)$ in the ansatz Eq.(\ref{ubt}), by requiring
that the thermodynamic potential is minimized : $\partial\Omega/\partial f(\zbf k)=0$.
Such a functional minimization leads to

\be
\tan 2f(\zbf k)=\frac{g I_q}{\bar \epsilon_q-\bar\nu}
\equiv\frac{\Delta}{\bar\xi_q}
\label{tan2f}
\ee
\noindent In the above, $\bar\epsilon_q=
(\epsilon(\kp)+\epsilon(\km))/2 =(\zbf k^2+\zbf q^2/4)/(2m)$ is the average kinetic energy of the two
condensing fermions. Similarly, $\bar \nu=(\nu_1+\nu_2)/2$, is
the average of the interacting chemical potential of the two fermions with
$\nu^a=\mu^a-g\rho^a$ and $\bar\xi_q=\bar\epsilon_q-\bar\nu$.
Further, we have defined $\Delta=-g I_q$,
with, $I_q$ as defined in Eq.(\ref{iq}). It is thus seen that the 
condensate function depends upon the 
{\em average} energy and the {\em average} chemical potential
of the fermions that condense.
Substituting  the solutions for the condensate functions given in Eq.
(\ref{tan2f}) in the expression for
$I_q$ in Eq.(\ref{iq}), we have the superfluid gap equation given by
\bearr
\Delta&=&-\frac{g}{(2\pi)^3}\int{d\zbf k}
\frac{\Delta}{2\bar\omega_q} \bigg[1-\sin^2\theta_1(\kp)
\nonumber\\&-&\sin^2\theta(\km)\bigg]
\label{del}
\eearr
In the above, 
 $\bar\omega_q=\sqrt{\Delta^2+\bar\xi_q^2}$ and the
 thermal functions $\theta^a(\zbf k,\beta)$ are still to be determined.

The minimization of the thermodynamic potential with respect to the
thermal functions $\theta_{a}(\zbf k)$ gives
\be
\sin^2\theta_a=\frac{1}{\exp(\beta\omega_a)+1},
\label{them}
\ee
with the quasi particle energies given as
$\omega_1(\kp)=\bar\omega_q+\delta_\xi$, and
 $\omega_2(\km)=\bar\omega_q-\delta_\xi$,
with $\delta_\xi\equiv(\xi_1-\xi_2)/2=
(\epsilon(\kp)-\epsilon(\km))/2-(\nu_1-\nu_2)/2
\equiv\delta_\epsilon-\delta_\nu$.
It is clear from the
dispersion relations that  
it is possible to have zero modes, i.e., $\omega^a=0$
depending upon the values of $\delta_\epsilon$
and $\delta_\nu$. So, although we shall have nonzero order
parameter $\Delta$, there can be fermionic zero modes, the so called 
gapless superconducting phase \cite {abrikosov,krischprl}. 

Now using these dispersion relations,
and the superfluid gap
equation (Eq.(\ref{del})), the thermodynamic potential 
(Eq.(\ref{Omega})) becomes
\bearr
\Omega&=&\frac{1}{(2\pi)^3}\int d\zbf k\bigg[(\bar\xi_q-\bar\omega)
\nonumber\\&-&
\frac{1}{\beta}\sum_a
\ln\left(1+\exp(-\beta\omega_a)\right)\bigg]-\frac{\Delta^2}{g}
\label{omgt}
\eearr
In what follows we shall concentrate on the case for zero temperature.
In that case the distribution functions, Eq.(\ref{them})
become $\Theta$ - functions
i.e. $\sin^2\theta^a=\Theta(-\omega_a)$. Further, using the identity
$\lim_{a\rightarrow \infty}\ln (1+\exp(-a x)/a=-x\Theta(-x)$, in 
Eq. (\ref{omgt}), the zero temperature thermodynamic potential
becomes
\bearr
\Omega&=&\frac{1}{(2\pi)^3}\int d\zbf k(\bar\xi_q-\bar\omega)\nonumber\\
&+&
\frac{1}{(2\pi)^3}\int d\zbf k\left[\omega_1\Theta(-\omega_1)+\omega_2
\Theta(-\omega_2)\right]
-\frac{\Delta^2}{g}.
\label{omega0}
\eearr

To compare the thermodynamic potential with respect to the normal matter,
we need to subtract out the zero gap and zero momentum ($\zbf q=\zbf 0$) 
part  from the thermodynamic potential given in Eq. (\ref{omega0}). 
This is given as
\bearr
\bar\Omega&\equiv&\Omega(\zbf q,\Delta,\delta)-\Omega(\zbf q=\zbf 0,\Delta=0,
\delta)\nonumber\\
&=&
\frac{1}{(2\pi)^3}\int d\zbf k\left(|\bar\xi_q|-\bar\omega\right)\nonumber\\
&+&
\frac{1}{(2\pi)^3}\int d\zbf k\left[\omega_1\Theta\left(-\omega_1\right)+\omega_2
\Theta\left(-\omega_2\right)\right]\nonumber\\
&-&\frac{1}{(2\pi)^3}\int d \zbf k (\xi-\delta_\nu)
\Theta(\delta_\nu-\xi)
-\frac{\Delta^2}{g},
\label{baromg}
\eearr
where, we have denoted $\xi=\zbf k^2/(2m)-\bar\nu$, the excitation 
energy with respect to the average fermi energy for normal matter.
Further, we have assumed, without loss of generality, $\delta_\nu> 0$. The 
superfluid gap in Eq.(\ref{baromg}) satisfies the gap equation
Eq.(\ref{del}) which at zero temperature reduces to
\bearr
\frac{1}{g}&=&-\frac{1}{(2\pi)^3}\int{d\zbf k}
\frac{1}{2\bar\omega_q} \bigg[1-\Theta \Big(-\omega_1(\kp)\Big)\nonumber\\
&-&\Theta \Big (-\omega_2 (\km)\Big)
\bigg].
\label{g}
\eearr
The above equation is ultraviolet divergent. The origin of this divergence
lies in the point like four fermion interaction which needs to be regularized.
 We tackle this problem by defining the regularized coupling in terms of 
the s-wave scattering length $a_s$ as was done in Ref. \cite{randeria,heiselberg} 
to access the strong coupling regime \cite{rapid,rupak} by subtracting out the
zero temperature and zero density contribution. The regularized gap equation
is then given as
\bearr
-\frac{1}{g_R}&\equiv&\frac{M}{4\pi a_s}\nonumber\\&=&\frac{1}{(2\pi)^3}
\int
d\zbf k\bigg[\frac{1-\Theta(-\omega_1)
\Theta(-\omega_2)}{2\bar\omega_q}\nonumber\\&-&
\frac{1}{2\epsilon(k)}\bigg]
\label{gr}
\eearr

Using Eq.(\ref{g})and Eq.(\ref{gr}) in Eq.(\ref{baromg}) to 
eliminate $g$ in favor of $g_R$, one can obtain
\bearr
\bar\Omega&=&
\frac{1}{(2\pi)^3}\int d\zbf k
\left(|\bar\xi_q|-\bar\omega_q+\frac{\Delta^2}{2\bar\omega_q}\right)\nonumber\\
&+&\frac{1}{(2\pi)^3}\int d\zbf k\left[(\omega_1
-\frac{\Delta^2}{2\bar\omega_q})
\Theta(-\omega_1)+
(\omega_2
-\frac{\Delta^2}{2\bar\omega_q})
\Theta(-\omega_2)\right]\nonumber\\
&-&\frac{1}{(2\pi)^3}
\int d \zbf k \bigg((|\bar\xi_q|+\delta_\xi)\Theta(-|\bar\xi_q|-\delta_\xi)\nonumber\\
&+&(|\bar\xi_q|-\delta_\xi)\Theta(-|\xi_q|+\delta_\xi)\bigg)d\zbf k,
\label{baromgq}
\eearr
where, $|\bar\xi_q|=(\zbf k^2+\zbf q^2/4)/2m-\bar\nu$ and $\delta_\xi=
\zbf k\cdot\zbf q/2m-\delta_\nu$.
 We might note here that the thermodynamic potential difference as
 in Eq.(\ref{baromgq}) is cut-off independent and is finite.
The only other quantities needed to calculate the thermodynamic potential
in Eq.(\ref{baromgq}) are the chemical potentials: the average chemical
potential of the two species $\bar\nu=(\nu_1+\nu_2)/2$
and their difference $\delta_\nu =(\bar\nu_1-\bar\nu_2)$. These two quantities 
can be fixed from the average number densities as
\bearr
&&\bar\rho=
\frac{1}{2}(\langle\psi_1^\dagger\psi_1\rangle+
 \langle\psi_2^\dagger\psi_2\rangle)\nonumber\\
&&=\frac{1}{2}
\frac{1}{(2\pi)^3}\int d\zbf k \left (1-\frac{\bar\xi_q}{\bar\omega_q}
\right) (1-\Theta(-\omega_1)-\Theta(-\omega_2)),
\label{rho}
\eearr

and the difference in number densities
\bearr
\delta_\rho&=&
\frac{1}{2}(\langle\psi_1^\dagger\psi_1\rangle-
 \langle\psi_2^\dagger\psi_2\rangle)\nonumber\\
&=&\frac{1}{2}\frac{1}{(2\pi)^3}\int d\zbf k\left(\Theta(-\omega_1)-
\Theta(-\omega_2)\right),
\label{drho}
\eearr
which does not depend on the condensate function explicitly.
In the following we shall discuss which phase is thermodynamically
favorable at what density as the chemical potential difference is varied
for a given coupling and  a given average density.

For numerical calculations, it is convenient to write the
Eq.s (\ref{gr}--\ref{drho})
in terms of dimensionless quantities. Thus we make the substitutions
$|\zbf k|=k_f x$, $\zbf q=k_f\zbf y$, $\Delta=\epsilon_f z$,
$\bar\nu=\epsilon_f \nu$, $\delta_\nu=\epsilon_f\delta$,
where, $k_f$ is the average fermi momentum and
$\epsilon_f$ is the corresponding fermi energy. In terms of these
dimensionless variables, the gap equation Eq.(\ref{gr}) reduces to

\bearr
-\frac{1}{8\pi k_fa}&=&\frac{1}{4\pi^2}\int_0^\infty dx x^2\left(
\frac{1}{\sqrt{(x^2+y^2/4-\nu)^2+z^2}}-\frac{1}{x^2}\right)\nonumber\\
&-&
\frac{1}{4\pi^2}\int_0^\infty dx x^2 \frac{1}{2\sqrt{(x^2-\nu)^2
+z^2}}\nonumber\\ 
&\times &\int_{-1}^{1} dt (\Theta(-\hwo) +\Theta(-\hwt)).
\label{gapd}
\eearr
Here, the quasi particle energies $\hat\omega_a$ ($a=1,2$) in units of 
average fermi energy are
\be
\hat \omega_1=\hat\omega_y+xyt-\delta ,\quad \omega_2=\hat\omega_y-xyt+\delta
\label{omg1}
\ee
where, $\hat\omega_y=\sqrt{(x^2+y^2/4-\nu)^2+z^2}$, where, $\nu$
 is the average of the excitation energies of the 
two quasi particles in units of fermi energy. Further, in Eq.(\ref{gapd}), 
the integration variable 
$t=\cos\theta $ and, we have chosen the momentum $\zbf q$ of the 
condensate to be in the z- direction.

Similarly, the average density equation Eq.(\ref{rho}), 
in units of $\k_f^3/6\pi^2$,
reduces to
\bearr
1&=&\frac{3}{2}\int_0^\infty dx x^2\int_{-1}^1 dt
 \frac{1}{2}\bigg[(1-\frac{\xi_y}{\hat\omega_y})\nonumber\\&+&
\frac{\xi_y}{\hat\omega_y}(\Theta(-\hwo)+\Theta(-\hwt))\bigg].
\label{rhod}
\eearr

Finally, the thermodynamic potential difference between the condensed
phase and the normal phase can be expressed in terms of these dimensionless 
variables as

\bearr
&&\bar\Omega=\frac{\epsilon_f\k_f^3}{2\pi^2}\bigg[\int x^2 dx
\left(|x^2-\nu+\frac{y^2}{4}|-
\hat\omega_y+\frac{z^2}{2\hat\omega_y}\right)\nonumber\\
&&+
\int x^2 dx\frac{dt}{2}\left((\hwo-\frac{z^2}{2\hat\omega_y})\Theta(-\hwo)
+(\hwt-\frac{z^2}{2\hat\omega_y})\Theta(-\hwt)\right)
\nonumber\\
&&-\frac{1}{2}\int x^2 dxdt
\bigg((|\xi_y|+\delta_{\xi_y} \theta(-\delta_{\xi_y}-\xi_y)\nonumber\\
&&+
(|\xi_y|-\delta_{\xi_y})\theta(\delta_{\xi_y}-\xi_y)\bigg)
\bigg]\nonumber\\
&&\equiv \frac{\epsilon_f\k_f^3}{2\pi^2}\bigg[I_{y1}+I_{y2}-I_{y3}\bigg],
\label{omgd}
\eearr
where, $\delta_{\xi_y}=xyt-\delta$ and $I_{yi}(i=1,2,3)$ are the three integrals in the above equation.
Eq.s (\ref{gapd}), (\ref{rhod}) and Eq.(\ref{omgd}) will form the basis for
our numerical investigation to decide regarding the phase structure.
Before going to the
numerical evaluations let us  discuss the different phases.
\section{Results}
\label{results}
\subsection{Homogeneous symmetric phase}
\label{homogeneous}
Let us first discuss the symmetric homogeneous  phase corresponding to
vanishing chemical potential difference ($\delta_\nu=0$). In that case
the renormalized gap equation, Eq.(\ref{gapd}), reduces to the usual
BCS gap equation
\cite{rapid,randeria}

\bearr
-\frac{1}{8\pi k_fa}&=&\frac{1}{4\pi^2}\int_0^\infty dx x^2\left(
\frac{1}{\sqrt{(x^2-\nu)^2+z^2}}-\frac{1}{x^2}\right)\nonumber\\
&=&I_z
\label{gap0}
\eearr

and, the number density equation, Eq.(\ref{rhod}), becomes
\bearr
1&=&\frac{3}{2}\int_0^\infty
dx x^2\left(1-\frac{x^2-\nu}{\sqrt{(x^2-\nu)^2+z^2}}\right)
\nonumber\\&=&
I_\rho
\label{rho0}
\eearr
where, $z$ and $\nu$ are respectively the superfluid gap and average chemical potential in units of fermi energy.

\begin{figure}
\resizebox{0.5\textwidth}{!}{%
  \includegraphics{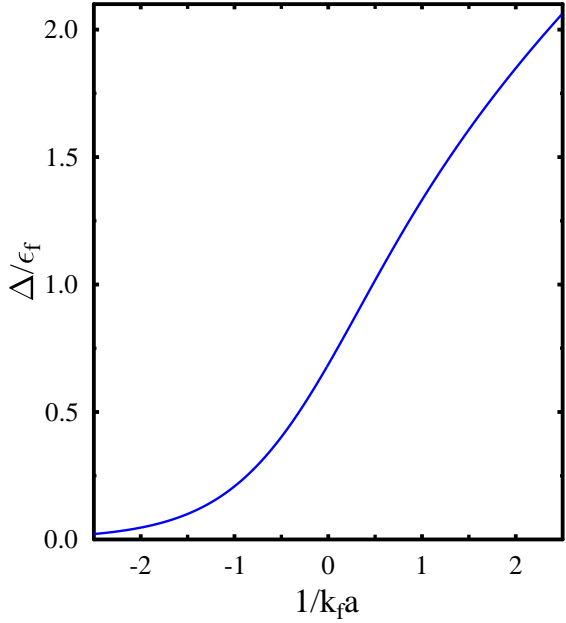}
}

\caption{Superfluid gap as a function of
coupling $1/k_fa$. The expoential rise of the gap 
in the BCS side ($k_fa < 0$) is as expected from
weak coupling results.}
\label{gapkfa}
\end{figure}

\begin{figure}
\resizebox{0.5\textwidth}{!}{%
  \includegraphics{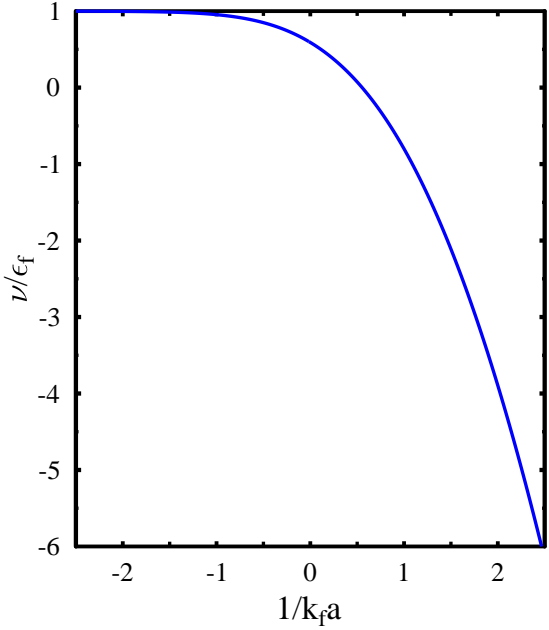}
}
\caption{Self consistent solution
for the chemical potential in units of fermi energy as a function of coupling
$1/k_fa$.}
\label{mukfa}
\end{figure}

It is easy to show, via integration by parts, that the integrals
$I_z$ and $I_\rho$ can be rewritten as

\be
I_z=\frac{1}{2\pi^2} \left(\nu I(\nu,z)-z^2 J(\nu,z)\right)
\label{iz}
\ee
and
\be
I_\rho=z^2
\left(I(\nu,z)+\nu J(\nu,z)\right)
\label{irh}
\ee
where, we have defined the integrals $I(\nu,z)$ and $J(\nu,z)$ as 
\be
I(\nu,z)=\frac{1}{2}\int_0^\infty \frac{dx}{\omega(x)}
\label{inuz}
\ee
and
\be
J(\nu,z)=\int_0^\infty \frac{x^2dx}{\omega(x)^3}
\label{jnuz}
\ee
with $\omega(x)=\sqrt{(x^2-\nu)^2+z^2}$

The nice thing about these integrals $I(\nu,z)$ and $J(\nu,z)$ is that
they can be expressed in terms of elliptic integrals \cite{marini}.
They can be written as
\be
I(\nu,z)=\frac{1}{2 (\nu^2+z^2)^{1/4}}F(\pi/2,\kappa)
\ee

and
\bearr
J(\nu,z)&=&
\frac{(z^2+\nu^2)^{1/4}}{z^2}E(\pi/2,\kappa)\nonumber\\&-&
\frac{F(\pi/2,k)}{2
(z^2+\nu^2)^{1/4}(\nu+\sqrt{z^2+\nu^2})}
\eearr
where, $F(\frac{\pi}{2},\kappa)$ and $E(\frac{\pi}{2},\kappa)$ are 
the elliptic integrals of first kind and second kind respectively. 
In the above,\\ $\kappa^2=(\nu+\sqrt{z^2+\nu^2})/(2\sqrt{z^2+\nu^2})$. 
The thermodynamic potential difference Eq.(\ref{omgd}) reduces to
\bearr
\bar\Omega(\delta=0, z)&=&
\frac{\epsilon_f k_f^3}{2\pi^2}
\int dx x^2\left(|x^2-\nu|-
\hat\omega+\frac{z^2}{2\hat\omega}\right)
\eearr
where, $\hat\omega=\sqrt{(x^2-\nu)^2+z^2}$.

In figures \ref{gapkfa} and  \ref{mukfa}
we plot the superfluid gap  and the chemical potential as  functions of
coupling $1/k_fa$ obtained by solving  the coupled gap equation (Eq.(\ref{gap0}))
 and number density equations
(Eq.(\ref{rho0})).
In the weak coupling BCS limit ($1/(k_f a)\rightarrow -\infty$),
we have $\nu>>z$ and the modulus parameter $\kappa^2 \simeq 1-z^2/(4\nu^2)$
is close to unity in the argument of the elliptic integrals.
Expanding the elliptic integrals in this limit, the gap equation
Eq.(\ref{iz}) reduces to
\be
\frac{1}{8\pi k_fa}=-\frac{\sqrt\nu}{4\pi^2}\left(\ln\frac{8\nu}{z}-2\right).
\label{weakgap}
\ee
This leads to the weak coupling BCS limit for the gap as
$z=8e^{-2}\exp(-\pi/(|k_fa|))$ in units of fermi energy.
Similarly in the BEC limit ($1/(k_f a)\rightarrow \infty$),
one expects tightly bound pairs with binding energy $E_b=1/(ma^2)$
and nondegenerate fermions with a large negative chemical potential
\cite{randeria}.
In this limit $\kappa^2\simeq z/(4\nu^2)<<1$. 
Taking this asymptotic limit of the elliptic integrals, the
chemical potential $\nu$ is half of the pair binding energy. The
corresponding gap is $z=(16/\pi)^{1/2}/\sqrt{k_fa}$ .

 As may be seen 
from Fig. \ref{mukfa}, the chemical potential changes sign at 
$\k_fa\sim 0.56$ signaling the onset of the BEC phase. Near to the 
unitary limit e.g. for $k_fa=0.01$, the ratio of condensate to the 
chemical potential  turns out to be $\Delta/\nu=1.195$, while the 
chemical potential itself is numerically evaluated to be  
$\nu/\epsilon_f=0.58$.
While the ratio of gap to the chemical potential
$\Delta/\nu$ is  very close to the value $\Delta/\nu=1.2$ obtained through
 quantum Monte Carlo simulations, the value of the ratio
of chemical potential to the fermi energy, $\nu/\epsilon_f$, is higher as
compared to the value  $\nu/\epsilon_f=0.42$ obtained from quantum Monte Carlo
simulation \cite{reddy}. Such results arising from the variational calculations
may be compared with the corresponding values obtained from
other nonperturbative calculations like $\epsilon$- expansion \cite{son} which  
turn out to be $\Delta/\nu\sim 1.31$ and $\nu/\epsilon_f=0.475$. We might 
note here that, a systematic analysis of the gaussian fluctuation around 
the saddle point approximation was considered in Ref. \cite{randeria}.
The corrections to the saddle point approximation were seen 
to be small for temperatures small compared to the
critical temperature, for {\em all} couplings \cite{randeria}.

\subsection{Isotropic superfluid with mismatch in chemical potentials}
\label{breached}
With a nonzero mismatch between chemical potentials i.e. $\delta_\nu\neq 0$,
let us consider superfluidity with a homogeneous condensate i.e. condensate 
with momentum  $\zbf q=\zbf 0$.
In that case the dispersion relations (in dimensionless variables)
of the quasi particles (Eq.(\ref{omg1}))
are given as
\be
\hat\omega_1=\sqrt{(x^2-\nu)^2+z^2}+\delta ; \quad
\hat\omega_2=\sqrt{(x^2-\nu)^2+z^2}-\delta. 
\label{omg120}
\ee
where, $\nu$, $z$ and $\delta$ are respectively the average chemical potential, the gap and the difference in chemical potentials in units of fermi energy.

For any value of the (dimensionless) gap $z>0$, the average chemical
potential $\nu$ and the difference in chemical potentials $\delta>0$,
the excitation branch
$\omega_1$ has no zeros similar to the case of a usual superconductor. The
quasi particles at the fermi surface have finite excitation energy
given by the gap  that is enhanced here by the mismatch $\delta$.
On the other hand, the  excitation branch $\omega_2$ can become
zero depending upon the value of  the gap $z$, the average chemical potential
$\nu$ and the mismatch parameter $\delta$.
Indeed, solving for the zeros for $\omega_2$ one obtains 
that $\omega_2$ vanishes at  momenta (in units of fermi momentum),
$x^2_{max/min}=\nu\pm\sqrt{\delta^2-z^2}$. Clearly, this has no solution
for $\delta < z$. Further, for the case of $\delta > z$, 
there will be no solutions if the average chemical potential $\nu$ is smaller than
$-\sqrt{\delta^2-z^2}$ since both $x_{max}^2$ and $x_{min}^2$ will become 
negative.  On the other hand, if $\delta >z$ and
$-\sqrt{\delta^2-z^2}<\bar\nu< \sqrt{\delta^2-z^2}$, then the expression
for $x_{max}^2$ is positive while that of $x_{min}^2$ is negative. Thus 
there is only
 one zero for $\omega_2$. This case will correspond to `interior gap' solution,
where, the unpaired fermions of the first species occupy the entire
effective fermi sphere. Finally, when $x^2_{max}> 0$ and $x^2_{min}>0$, there
are two zeros for $\omega_2$. This will correspond to breached pairing
case where unpaired fermions of first species occupy the states between
the two fermi spheres decided by $x_{max}$ and $x_{min}$.

With nonzero mismatch in the chemical potential, the gap equation for the
homogeneous phase is given as
\bearr
-\frac{1}{8\pi k_fa}&=&\frac{1}{4\pi^2}\int dx x^2\left(
\frac{1}{\bar\omega(x)}-\frac{1}{x^2}\right)\nonumber\\&-&
\frac{1}{4\pi^2}\int_{x_{min}}^{x_{max}}
 x^2 \frac{1}{\bar\omega(x)}dx\nonumber\\
&\equiv& I_z-I_\delta
\label{hgapd}
\eearr
Here, we have introduced the notation
$\bar\omega(x)=$$\sqrt{(x^2-\nu)^2+z^2}$ as the average of the
quasi particle energies of the two species with $\nu$ as the average 
chemical potential in units of fermi energy.

Further, the thermodynamic potential of Eq. (\ref{omgd})
reduces for homogeneous condensates to,
\bearr
&&\bar\Omega=\frac{\epsilon_f\k_f^3}{2\pi^2}
\bigg[\int x^2 dx
\left(|x^2-\nu|-
\bar\omega(x)+\frac{z^2}{2\bar\omega}\right)\nonumber\\
&&+\int x^2 dx\left((\omega_1-\frac{z^2}{2\bar\omega(x)})\Theta(-\omega_1)
\right)
\nonumber\\
&&-\int x^2 dx \Theta(\delta-|x^2-\nu|)(|x^2-\nu|-\delta).
\bigg]\nonumber\\
&&\equiv\frac{\epsilon_f k_f^3}{2\pi^2} (I_1+I_2-I_3),
\label{homega}
\eearr
where, we have defined the three integrals appearing in Eq.(\ref{homega})
as $I_1$, $I_2$ and $I_3$ respectively.
 The integral $I_3$ can be evaluated directly as
\bearr
I_3&=&
\int x^2 dx \Theta(\delta-|x^2-\nu|)(|x^2-\nu|-\delta)\nonumber\\
&=&\frac{4}{15}\nu^{5/2}-\frac{2}{15}(\Theta(\nu_+)\nu_+^{5/2}+
\Theta(\nu_-)\nu_-^{5/2})
\label{i3}
\eearr
with $\nu_\pm=\nu\pm\delta$.

Similarly, the integral $I_2$ can be rewritten as
\bearr
I_2&=& \int x^2 dx(\omega_1-\frac{z^2}{2\bar\omega(x)})
\Theta(-\omega_1)\nonumber\\
&=&\int_{x_{min}}^{x_{max}} x^2 dx \left(\bar\omega(x)-\delta-
\frac{z^2}{2\bar\omega(x)}\right)
\eearr

\begin{figure}
\resizebox{0.5\textwidth}{!}{%
  \includegraphics{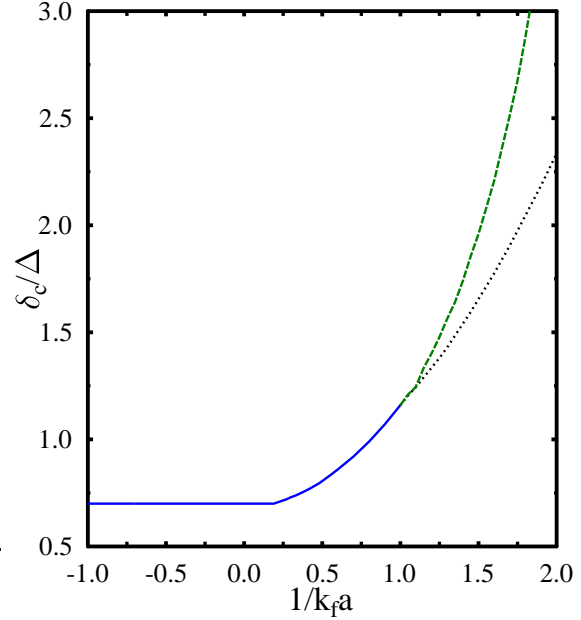}
}
\caption{\em{ 
 Ratio of critical chemical potential difference
to the gap as a function
of coupling strength $1/k_fa$. Gapless phase appears for
coupling $1/k_fa > 1$. The region between the dotted
and the dashed line shows the gapless phase.
}}
\label{kfadel}
\end{figure}

Before going to the numerical solution for the homogeneous condensates with
mismatch in chemical potential, let us discuss the weak coupling BCS
case. This limit is characterized by the average chemical potential being equal to the fermi energy so that
$\nu=1$ and is  much larger than the magnitude of both the chemical 
potential mismatch 
$\delta$ in dimensionless units and 
the gap $z$, also in dimensionless units. Thus
this weak coupling analysis excludes the scenario with  single fermi surface
i.e. with the excitation energy $\omega_2$ having a single zero.
Further, we here compare the thermodynamical potential for fixed
chemical potentials. It is expected that the same analysis with
fixed number densities will lead to different results -- namely
a mixed phase scenario which is an inhomogeneous phase
where a fraction of the space is in normal phase while the remaining is in the BCS
phase \cite{rupak}.
In this weak coupling limit, we can have ordinary superconductivity
without gapless modes or can have breached pairing with $\omega_2$
becoming zero at two values of the momentum. The latter situation will arise 
for much larger values of the mismatch parameter $\delta$ as compared to 
the gap $z$. 
In the weak coupling limit, the integral $I_1$ becomes
\be
I_1=\int x^2 dx \left(|x^2-\nu|-
\bar\omega(x)+\frac{z^2}{2\bar\omega}\right)\simeq -\frac{\sqrt{\nu}z^2}{4}
\label{i1bp}
\ee
In the same limit, $x_{min}$=$\sqrt{\nu}=x_{max}$, so that, $I_2\simeq 0$.
Finally, the integral $I_3\simeq \sqrt{\nu}\delta^2/4$. Thus in the weak
coupling limit, the thermodynamic potential difference between the condensed
 and the normal phase is
\be
\bar\Omega\simeq\frac{\epsilon_fk_f^3}{2\pi^2}(2\delta^2-z^2)
\ee
For the ratio of the chemical potential difference to the superfluid gap
$\delta/z<1/\sqrt{2}$, the Clogston--Chandrasekhar
 limit \cite{clog}, the thermodynamic potential difference becomes negative. 
In this regime of $\delta/z$, the gap is the same 
as the $\delta=0$ gap and 
the density remains the same even though $\delta$ is nonzero. This critical chemical potential difference below which BCS pairing is possible depnds on the coupling strength.

In Fig.\ref{kfadel} we plot, as a function
of the coupling $1/\k_fa$, the quantity
$\delta_{c}/\Delta$ the ratio of the maximum
chemical potential difference to the gap that can sustain pairing.
For the region below the solid and the dotted line corresponds to
the parameter region where BCS pairing is the state of minimum thermodynamic
potential. As may be seen from the figure, for weak
coupling, the critical chmecal potential
difference (in units of superfluid gap) 
approaches the Clogston--Chandrasekhar limit. As coupling
is increased from BCS to BEC side $\delta_c/\Delta$ increases monotonically.

We do not find any breached pairing solutions in the BCS region with lower
thermodynamic potential as compared to the normal matter. However, we do find 
the interior gap solution in a
range of $\delta_\nu$ in the strong coupling BEC regime with 
negative average chemical potential. This happens for $1/k_fa>1$. This is given by the region between the dotted and the dashed line in Fig.\ref{kfadel}. In this region the density difference between the two species is also nonzero.
In the region of the parameter space that is above the solid line
or to the left of the dashed line is the region where no pairing is possible 
and the normal matter is the state of minimum thermodynamic potential.

\begin{figure}[htbp]
\begin{center}
\resizebox{0.5\textwidth}{!}{%
  \includegraphics{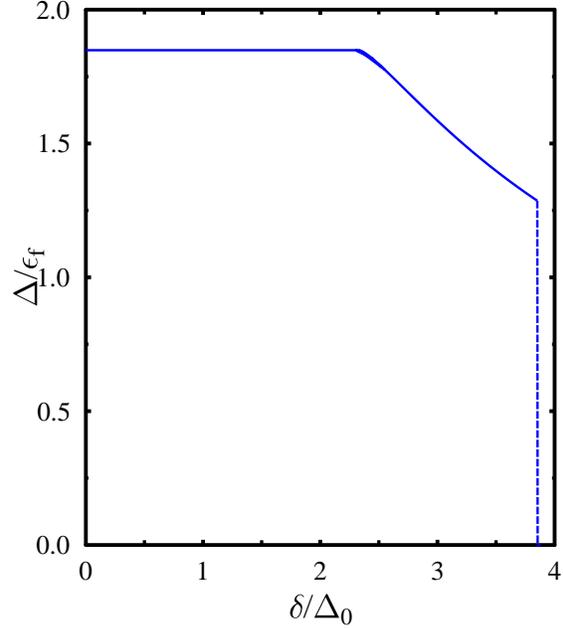}
}
\end{center}
\caption{Superfluid gap as a function of 
the difference in the chemical potentials of the two species.
The curve is plotted for $1/k_fa=2$.
}
\label{Deldel}
\end{figure}

In Fig.\ref{Deldel} we plot the gap as a function of difference 
in chemical potential
for $1/k_fa=2$ as a typical value for 
strong coupling BEC regime where gapless phase exists.
In this region, the thermodynamic potential difference 
between the condensed phase and the normal phase is negative. 
The value of the gap decreases continuously
from the symmetric BCS phase to the gapless phase as $\delta_\nu$ is increased
leading to nonzero difference in the number densites. The transition from gapless
phase to the normal phase which occurs at $\delta_\nu\simeq 3.85\Delta_0$ 
is however discontinuous. In the gapless phase the density difference between
the two species is non vanishing.
\begin{figure}
\resizebox{0.5\textwidth}{!}{%
  \includegraphics{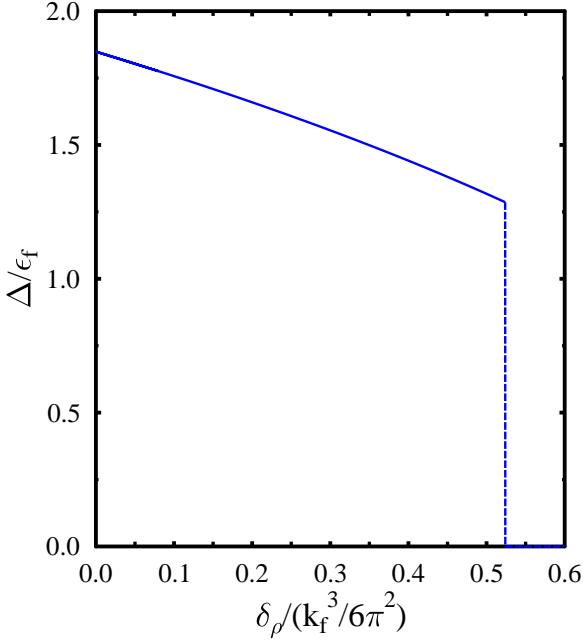}
}
\caption{\em{ Superfluid gap as a function of 
difference in number densities of the condensing
species.  The curve is plotted for $1/k_fa=2$.
}}
\label{rhdel}
\end{figure}

\begin{figure}
\resizebox{0.5\textwidth}{!}{%
  \includegraphics{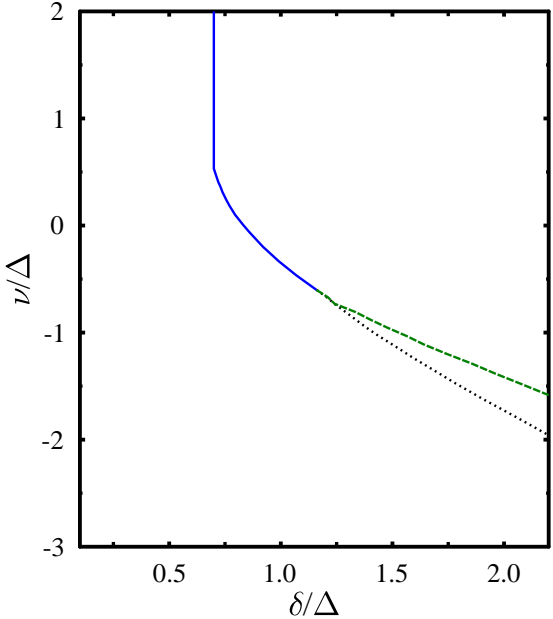}
}
\caption{\em{ 
The ratio of average chemical potential to the gap
as a function of difference 
of chemical potential 
The region between the dotted
and the solid line shows the gapless phase.
}}
\label{phdel}
\end{figure}

As mentioned earlier, in our calculations we do not keep the difference in the densities of the two species fixed. The calculations are performed with a fixed average density and given value of chemical potential difference. In the gapless phase, the density difference comes out  to be nonvanishing.
In Fig.\ref{rhdel} the dependence of the gap on the density difference 
$\delta_\rho$
is shown for the BEC region where gapless phase exists. We have taken here
the value of $1/\k_fa$ as equal to 2. Superconductivity is supported 
for a maximum density difference of $\delta_\rho\simeq 0.52$ in units of 
$k_f^3/6\pi^2$
beyond which the system goes over to the normal matter with zero gap. 
For coupling $1/k_fa< 1$, we do not find any superfluid phase free energetically
favorable for any non zero value for the density difference, although
a chemical potential difference can still support a Cooper paired  BCS phase
with zero net momentum..

In Fig.\ref{phdel}, we have plotted the phase diagram in the plane of
average chemical potential and the chemical potential difference, both
the quantities being normalized with respect to the superfluid gap. The 
region above and to the right of the solid line correspond to positive 
thermodynamic potential and is not stable.
The region below  and to the left of the solid line shows Cooper
pairing. The region between the dashed and the solid line shows the
gapless phase.
The vertical line at $\delta/\Delta=1/\sqrt{2}=0.717$ indicates the Clogston
Chandrasekhar limit.  These results, corresponding to 
explicit solutions for the gap as 
a function of coupling are in conformity with the results obtained
in Ref. \cite{andreas}.

\subsection{The LOFF Phase: anisotropic superfluid with $\zbf q\neq 0$}
\label{loffphase}
When the value of  the difference in chemical potentials, $\delta$,
 exceeds the Clogston-Chandrasekhar limit,
it can be free energetically favorable to have condensates with nonzero
momentum $\zbf q$, given as in Eq.(\ref{iq}). The dispersion relations for the
quasi particles are given as
\bearr
\omega_1(\zbf k)&=&\bar\omega(\zbf k)+\delta_\xi\nonumber\\&=&
\sqrt{\left(
\left(\zbf k^2+\frac{\zbf q^2}{4}\right)/2m+\bar\nu\right)^2+\Delta^2}
\nonumber\\&-&
\frac{\zbf k\cdot\zbf q}{2m}-\delta_\nu
\eearr
\bearr
\omega_2(\zbf k)&=&\bar\omega(\zbf k)-\delta_\xi\nonumber\\
&=&\sqrt{\left(
\left(\zbf k^2+\frac{\zbf q^2}{4}\right)/2m-\bar\nu\right)^2+\Delta^2}
\nonumber\\&+&
\frac{\zbf k\cdot\zbf q}{2m}+\delta_\nu
\eearr
The corresponding 
gap equation, number density equation and the thermodynamic potential 
are already given in Eq.s (\ref{gapd}), (\ref{rhod}) and (\ref{omgd})
 respectively.

As in the previous subsection, before going to discuss the detailed numerical
solutions for strong couplings, let us consider the weak coupling 
limits of the above equations which can give an insight to the numerical 
solutions that can be obtained in the appropriate limit.
In this limit, the average chemical potential is same as the
fermi energy sothat$\sqrt{\nu}=\bar\nu/\epsilon_f\sim 1 $ and is much larger
than the momentum (in units of fermi momentum) $y$,
gap  ( in units of fermi energy) $z$ and the asymmetry in chemical potential
(in units of fermi energy) , $\delta$.
Further, the excitation energies in units of fermi energy can be approximated as
\be
\omega_{1,2}=\bar\omega_y\pm\delta_t
\ee
where, $\bar\omega_y\simeq \sqrt{z^2+4\nu(x-\sqrt{\nu_y})^2}$,
and, $\delta_t\simeq yt-\delta=\delta(\rho t-1)$, with $\rho=y/\delta$. Just to remind ourselves, $y$ is the condensate momentum in units of $k_f$ and 
$\delta$ is half of the difference in the chemical potentials of the 
two species in units of fermi energy $\epsilon_f=k_f^2/2m$. Using Eq.(\ref{weakgap}),
one can eliminate the coupling $1/k_fa$ from the gap Eq.(\ref{gapd})
to obtain the weak coupling LOFF gap equation as
\bearr
&&\ln\left(\frac{z_0}{z}\right)\nonumber\\&&
=\frac{\sqrt{\nu}}{2}\int dx dt\frac{1}{\bar\omega_y(x)}
\left(\Theta(-\omega_1)+\Theta(-\omega_2)\right).
\label{gapwe}
\eearr
The theta functions in the integral above restrict the limits of
$x$ integration to be in the range $(x_{min},x_{max})$,with,
\be
x_{max/min}=\sqrt{\nu}\pm\frac{1}{2\sqrt{\nu}}(\delta_t^2-z^2)^{1/2}
\ee
Performing the $x$-integration within the limit, the gap equation,
 Eq.(\ref{gapwe}), reduces to
\bearr
&&\ln\left(\frac{z_0}{z}\right)\nonumber\\&&=\frac{1}{4}\int  dt \Theta(|\delta_t|-z)
\ln\left(\frac{|\delta_t|+\sqrt{\delta_t^2-z^2}}
{|\delta_t|-\sqrt{\delta_t^2-z^2}}\right).
\label{gapwe1}
\eearr
Again, the theta function  restricts the angular integration where either
or both the quasi particle excitation  energies ($\omega_{1,2}$) become gapless. 
This occurs for the case when the gap $z$ is less than $(1+\rho)\delta$
as otherwise the domain of integration specified by the 
theta function $\Theta(|\delta_t|-z)$ becomes null. Depending upon
 the value of the gap $z$ as compared to $(\rho-1)$, either one of the
two quasi particles or both become gapless \cite{ren}.
Using this fact, the integration over $t$ can be performed  in 
Eq.(\ref{gapwe1}) leading to the weak coupling  LOFF gap equation as
\bearr
&&\ln\left(\frac{z_0}{z}\right)\nonumber\\&&+\left[
\frac{\rho+1}{4\rho}\left(\ln\frac{1-x_1}{1+x_1} +2x_1\right)
+\frac{\rho-1}{4\rho}\left(\ln\frac{1-x_2}{1+x_2} +2x_2\right)
\right]\nonumber\\&&=0,
\label{gapren}
\eearr
where, $x_1$, $x_2$ are parameters
\be
x_1=\Theta\left(1-\frac{z}{(\rho+1)\delta}\right)\left(1-
\frac{z^2}{\delta^2(\rho+1)^2}\right)^{1/2}
\label{x1}
\ee
and,
\be
x_2=\Theta\left(1-\frac{z}{|\rho-1|\delta}\right)\left(1-
\frac{z^2}{\delta^2(\rho-1)^2}\right)^{1/2}
\label{x2}
\ee
We note here that the gap equation Eq.(\ref{gapren}), is identical to the
one derived in Ref.\cite{takada} and Ref.\cite{ren} for LOFF phase in
the superconducting alloy with paramagnetic impurities and 
in quark matter respectively.

\begin{figure}
\resizebox{0.5\textwidth}{!}{%
  \includegraphics{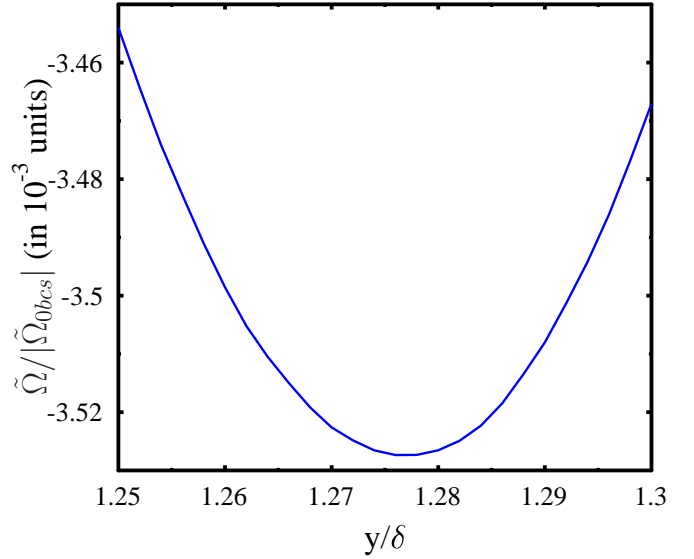}
}
\caption{\em{ The  thermodynamic potential between the LOFF and
normal state as a function of LOFF momentum. The thermodynamic
potential is normalized to the magnitude of the thermodynamic potential of 
the BCS state at $\delta=0$. The momentum is normalized with respect to
the chemical potential difference. This curve corresponds to the value of
the coupling $1/k_fa=-1.5$ and $\delta_\nu=0.72\Delta_0$.}}
\label{poty}
\end{figure}
 Next, let us look at the thermodynamic potential Eq.(\ref{omgd}) in the
weak coupling limit. The thermodynamic potential in Eq.(\ref{omgd})
can be written as
\be
\bar\Omega=\frac{\epsilon_fk_f^3}{2\pi^2}\left(I_{y1}+I_{y2}-I_{y3}\right)
\label{omegawe}
\ee
where, $I_{yi}$($i$=1,2,3) are the three integrals of Eq.(\ref{omgd}).
It is easy to show, using Eq.(\ref{i1bp}), that in the weak coupling
limit
\be
I_{y1}\simeq -\frac{\sqrt{(\nu-y^2/4)}z^2}{4}
\label{iy1}
\ee
The evaluation of the integral $I_{y2}$, is similar to the evaluation
of the integral on RHS of the gap equation  Eq.(\ref{gapwe}). The result is
\be
I_{y2}\simeq -\frac{\sqrt{\nu}\delta^2}{12}
\left[\frac{(\rho+1)^3}{\rho}x_1^3+\frac{(\rho-1)^3}{\rho}x_2^3
\right].
\label{iy2}
\ee
The integral $I_{y3}$ is straightforward to evaluate and is
given as
\bearr
I_{y3}&=&\frac{4}{15}(\bar\nu-\frac{y^2}{4})^{5/2}-\frac{2}{15}\left(
(\nu+\delta)^{5/2}+(\nu-\delta)^{5/2}\right)\nonumber\\&\simeq&
-\frac{\sqrt\nu}{2}(
\nu \frac{y^2}{3}+\delta^2)
\eearr
Collecting all the terms, the thermodynamic potential Eq.(\ref{omegawe})
is given as
\bearr
\hat\Omega &\simeq &
 \frac{\epsilon_fk_f^3}{2\pi^2}\frac{\sqrt\nu}{2}
\bigg[
\delta^2+\frac{y^2}{3}\nu-\frac{z^2}{2}\nonumber\\&-&\frac{\delta^2}{6\rho}
\left((\rho+1)^3x_1^3+(\rho-1)^3x_2^3\right)
\bigg].
\label{omgren}
\eearr
 It is reassuring to note that the expression Eq.(\ref{omgren}) for 
the thermodynamic potential is same as in Ref.\cite{takada}. This is
also identical to the thermodynamic potential considered for quark
matter in LOFF phase in Ref.\cite{ren} when degeneracy factors for the color 
and flavor are taken into account.

Similar to Ref. \cite{ren}, it can be shown that, for the case of
a small gap as compared to the chemical potential mismatch parameter, 
i.e. $z<<\delta$,
which will be the case near a second order phase transition to normal
matter, the gap is given as \cite{takada,ren}
\be
z^2=4(\rho_c^2-1)\left(1-\frac{\delta}{\delta_c}\right)
\ee
and, the momentum $\rho=|\zbf q|/\delta_\nu$ is given as \cite{takada,ren}
\be
\rho=\rho_c+\frac{1}{4}\frac{\rho_c}{\rho_c^2-1}z^2.
\ee
In the above, $\rho_c$ satisfies the equation
\be
\frac{1}{2\rho_c}\ln\frac{\rho_c+1}{\rho_c-1}=1
\ee
and 
\be
\frac{\delta_c}{z_0}=\frac{1}{2\sqrt{\rho_c^2-1}}\simeq 0.754,
\ee
for $\rho_c \simeq 1.2$.
\begin{figure}
\resizebox{0.5\textwidth}{!}{%
  \includegraphics{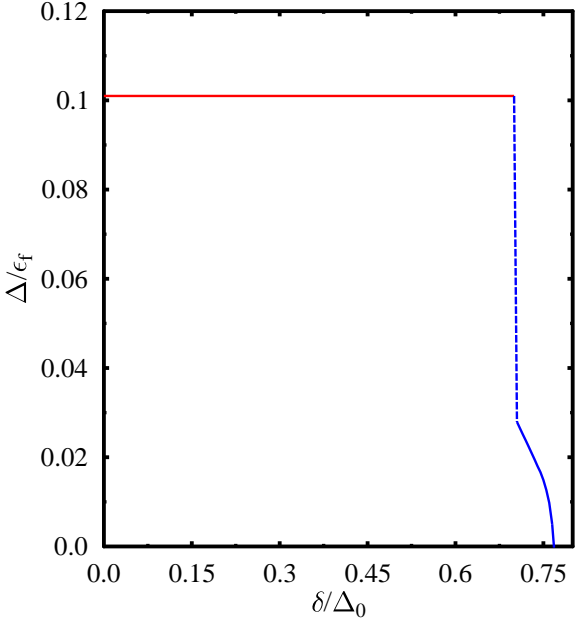}
}
\begin{center}
\end{center}
\caption{\em{ Superfluid gap as a function of chemical potential
difference. The coupling parameter here is taken as $1/k_fa=-1.5$.
}}
\label{gapdel}
\end{figure}

Here, $\delta_c=0.754 z_0$ is the maximum value of the chemical potential
difference that can support the LOFF phase beyond which the
system goes to the normal matter phase. For a general gap parameter
which need not be small as compared to the chemical potential difference,
one has to solve the gap equation Eq.(\ref{gapren}) numerically. This is done
for a given value of  LOFF momentum $\rho$ so that the thermodynamic 
potential as given in Eq.(\ref{omgren}) is calculated for a given value
of the momentum. The minimization of the thermodynamic potential numerically
over the magnitude of momentum gives the best value of the LOFF momentum.

In the present calculations, however,
we do not use the weak coupling gap equation, Eq.(\ref{gapren}). Instead,
for given values of  coupling $1/k_fa$, the chemical potential
difference $\delta_\nu$ and LOFF momentum $q$, the coupled equations
Eq.(\ref{gapd}) and Eq.(\ref{omgd}) are solved in a self consistent manner.
Using the values of the gap and the average chemical potential
 so obtained we calculate the thermodynamic potential using Eq.(\ref{omgd}).
 For the numerical analysis, we
first start with the weak coupling BCS regime $\delta_\nu=\delta_{max}$, the boundary
that separates the LOFF phase and the normal phase. For $\delta_\nu< \delta_{max}$
, we solve the gap Eq. (\ref{gapd}) for a given value of $|\zbf q|$. Solution
to the LOFF gap equations exist for a range of $|\zbf q|$. For each value 
of $\delta$ we can determine the `best $|\zbf q|$': the choice of $|\zbf q|$ 
that has the lowest thermodynamic potential.
In Fig.\ref{poty}, the difference of
thermodynamic potential between the LOFF phase and the normal matter
is plotted as a function of the LOFF momentum for coupling $1/k_fa=-1.5$
and the chemical potential difference $\delta_\nu=0.72\Delta_0$. We have 
normalized the thermodynamic potential with respect to the magnitude 
of the thermodynamic potential for the BCS phase. 
We note here that at this value of
$\delta$, which is greater than the Clogston--Chandrasekhar
limit, the normal matter has lower thermodynamic potential than the
homogeneous BCS phase.
Thus the LOFF state is  favored  when the thermodynamic potential
difference that is plotted in Fig.\ref{poty} is negative. For weak coupling
this happens when $\delta_\nu > 0.705\Delta_0$, a value slightly lower than the 
Clogston--Chandrasekhar limit. In this region there are also other solutions
of gap equation with homogeneous gap, but with higher value of the thermodynamic
potential. At each $\delta < \delta_{max}$ we also compare the
 thermodynamic potential for the `best $|\zbf q|$' and that of homogeneous BCS phase. 
We see that
for $\delta > \delta_{min}$, LOFF state has lower thermodynamic potential as
compared to
the homogeneous BCS phase. At $\delta_{min}$ there is a first order phase 
transition between
the LOFF and the BCS phases. Thus in the `LOFF window' $\delta_{\max}-\delta_{min}$,
 the LOFF state has lower energy as compared to the BCS or the normal phase. 
 For weak coupling,
our numerical values for the condensate, the momentum of the condensate
and the LOFF window match with those obtained in Ref.\cite{loff,takada}.
\begin{figure}
\resizebox{0.5\textwidth}{!}{%
  \includegraphics{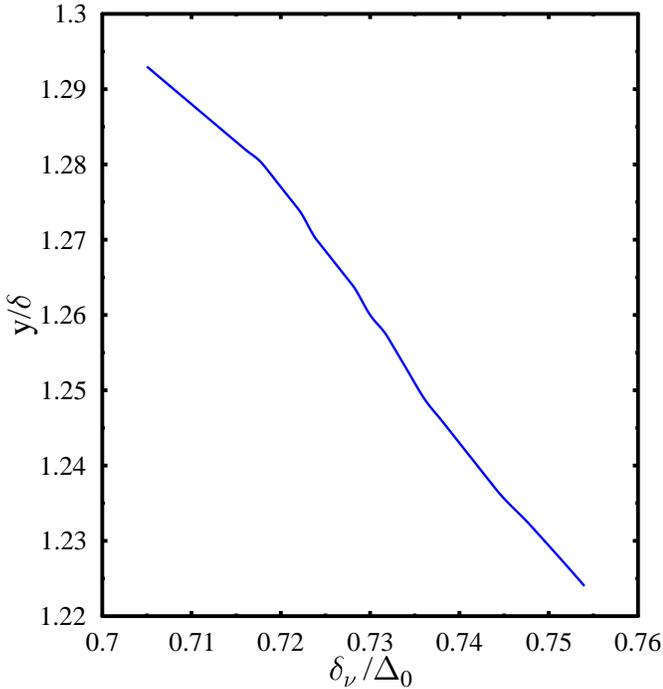}
}
\caption{\em{  LOFF momentum in units of mismatch in chemical potential
as a function of mismatch in chemical
potential in units of gap at zero mismatch $\Delta_0$. The coupling here 
is $1/k_fa=-1.5.$
}}
\label{ydel}
\end{figure}

 The variation of the gap
parameter as a function of difference in chemical potential is shown in Fig.
\ref{gapdel} for coupling $1/k_fa=-1.5$. Beyond $\delta_{max}$, the stress
 of difference in the chemical potentials
leads to the vanishing of the condensate, as a second order phase transition. 
The variation of the LOFF momentum with respect to the chemical potential 
difference is shown in Fig.\ref{ydel}. With $y=|\zbf q|/k_f$, ($\zbf q$,
being the LOFF momentum), we have plotted here the ratio $y/\delta$ as
a function of the chemical potential difference $\delta$ which decreases
monotonically from a value 1.29 at $\delta_\nu/\Delta_0=0.704$ to about 1.22
at $\delta_\nu/\Delta_0=0.754$. These values are similar to those
obtained in Ref.\cite{takada,ren} in the weak coupling limit.

\begin{figure}
\resizebox{0.5\textwidth}{!}{%
  \includegraphics{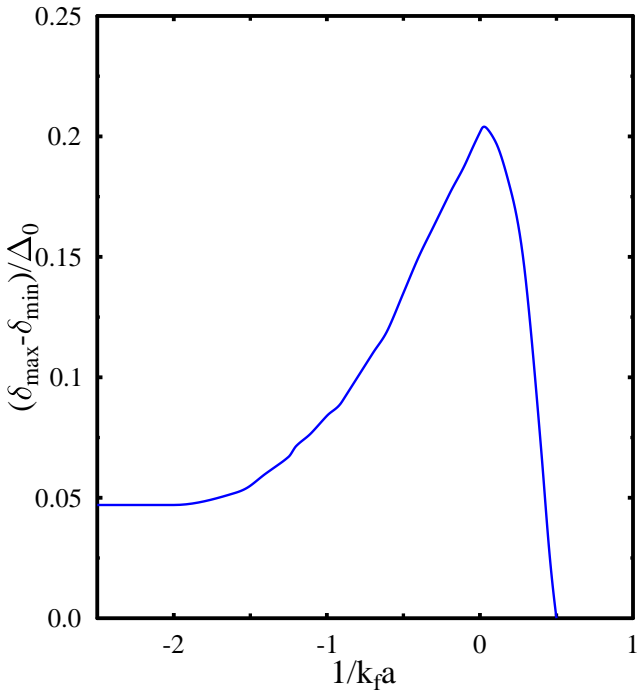}
}
\caption{\em{ The LOFF window in units of
the gap at zero temperature as a function of coupling $1/k_fa$}}
\label{winloff}
\end{figure}
Next we explore how the width of the LOFF window depends upon the strength of the 
coupling.
 We see that from the weak coupling BCS side, as we increase the
coupling the lower limit $\delta_{min}$ of the LOFF window 
decreases very slowly while the upper limit $\delta_{max}$ increases.
We might note here that the thermodynamic potential difference between the
LOFF state and the normal matter is very small. We had seen in the last
subsection that the Clogston -- Chandrasekhar limit  remains almost constant in the
BCS side of the Feshbach resonance (see e.g. Fig.\ref{kfadel}). 
$\delta_{min}$  which is very near to this value, thus
is almost constant on the BCS side of Feshbach resonance. With $\delta_{max}$
increasing in this region, the LOFF window increases. However, the 
Clogston -- Chandrasekhar limit increases with the coupling  for positive 
scattering length and $\delta_{min}$ of the LOFF window also increases resulting
in reduction of the LOFF window in the positive coupling region. 
Thus the LOFF window turns out to be a
non monotonic function of the coupling  and it becomes maximum at 
$1/k_fa\simeq0.04$.  Beyond this it decreases rapidly. For coupling greater 
than  $1/k_fa=0.5$; there is no longer any window of mismatch in chemical potential
in which LOFF state occurs. However, these results can be modified with a
more involved ansatz of multiple plane waves as opposed to the single
plane wave  ansatz as considered in Eq.(\ref{u1}) or Eq.(\ref{iq}). Such multi plane wave
ansatz was suggested to be favorable near the upper edge of the
LOFF window \cite{crystal}. In this context, it is worthwhile to mention also
that the LOFF window can be considerably expanded if one considers optical 
lattices\cite{koponen} rather than the pairing in free 
space as considered here.
As discussed earlier, for couplings $1/k_fa > 1$, the
the interior gap state is preferred within a certain range of mismatch in 
chemical potential.

\section{Summary and discussions}
\label{summary}
We have considered here a variational approach to discuss the ground state 
structure for a system of fermionic atoms with mismatch in their fermi momenta.
An explicit construct for the ground state is considered to describe two
fermion condensate with nonzero momentum. The ansatz functions including the
thermal distribution functions have been determined from the minimization 
of the thermodynamic potential. This is done for fixed chemical potentials. 
A similar analysis for fixed number densities leads to different results 
\cite{rupak}.

For comparison of thermodynamic potential, we have subtracted out the free 
energy of normal phase (with $\Delta=0$ and $\zbf q =\zbf 0$) as in 
Eq.(\ref{Omega}). This difference is finite without introducing any 
arbitrary cut off in the momentum \cite{zhuang}. The four fermi coupling
is also renormalized as in Eq.(\ref{gr}), by subtracting out the vacuum 
contribution and relating it to the s-wave scattering
length as in Ref.s \cite{rapid,randeria,heiselberg}. 
This makes all the quantities well defined and finite. 
Rewriting the gap equations and the number density equations
for the symmetric case in terms of elliptic integrals as in Eqs. (\ref{iz})
and (\ref{irh}) is particularly useful regarding the numerical evaluation. 

We have not calculated here the Meissner masses \cite{rischkeshov,ren} or 
the number susceptibility \cite{andreas} to discuss stability of different
phases by ruling out regions of the parameter space for the gap function
and the difference in chemical potentials. Instead, we have solved the
gap equations and the number density equations self consistently and 
have compared the thermodynamic potentials. This apart, we have also 
extended  the analysis to incorporate the possibility of condensate 
with finite momentum through the the ansatz for the ground state. 
In certain regions of chemical potential difference and couplings 
we have  multiple solutions for the gap equation. In such cases 
we have chosen the solution which has the least
value for the thermodynamic potential.

In the weak coupling BCS limit we obtain the usual LOFF solution 
i.e. the condensate with finite momentum, in a small window of the chemical 
potential difference of about 0.05 times the gap at zero chemical 
potential difference. This LOFF window increases as the coupling is increased 
and becomes maximum at $1/k_fa\sim0.04$ and vanishes at $1/k_fa=0.5$. 
Within the present approach we do not see any breached pairing solution 
being favored on the BCS side of the Feshbach resonance. However,
interior gap superfluid phase with a single fermi surface appears to be possible
on the strong coupling BEC side of the Feshbach resonance
with negative chemical potential similar to results in Ref.\cite{andreas}. 
The transition between BCS to LOFF phase is a first order one with the order 
parameter varying discontinuously at $\delta_{min}$, while the LOFF phase 
to normal phase transition is second order at chemical potential difference 
$\delta_{max}$. On the other hand, in the strong coupling BEC region, 
the phase transition from BCS to interior gap phase is a second order 
phase transition while the phase transition from interior gap
phase to the normal phase is a first order phase transition as the 
chemical potential difference is varied.  

These results are of course limited by the simplified ansatz
as considered here giving a unified description for the uniform as well
as spatially modulated superfluid. Similar ansatz  with  uniform condensates
has been used earlier interpolating the  BCS 
($g \rightarrow 0^+$, $1/a\rightarrow -\infty$) and 
BEC  ($g \rightarrow \infty$, $1/a\rightarrow \infty$) limits  for the case of
zero chemical potential difference \cite{randeria,marini}. The results 
as obtained here might nevertheless be regarded as a reference solution with
which other numerical or analytical solutions obtained from more involved ansatz
for the ground state can be compared. 

Although the present numerical analysis has been done for zero temperature, 
the expressions to include temperature effects have been derived here. 
Clearly, the effect of fluctuations involving the corrections from the
collective modes will play an important role for high temperatures
particularly for the strong coupling \cite{randeria}.
The effect of including different masses of the two species as well as
using a realistic potential for the two atomic species will be interesting 
regarding the study of the phase structure. Further, the study of
some of the transport properties in different phases including viscosity 
would be very interesting for cold atomic superfluids. Some of these
calculations are in progress and will be reported elsewhere \cite{amppn}.

\section{Acknowledgements}
We would like to thank B. Deb, P.K. Panigrahi, A. Vudaygiri, 
D. Angom, S. Silotri, B. Chatterjee for many illuminating discussions.
HM would like to thank organisers of the meeting on 'Interface of
QGP and Cold atom physics ' at ECT$^{*}$, Trento and acknowledges
discussions with D.Rischke, Y. Nishida, G.C. Strinati,
P. Piere and C. Lobo  during the meeting.
AM would  like to acknowledege financial support from Department of Science
and Technology, Government of India (project no. SR/S2/HEP-21/2006).

\def\loff{A.I. Larkin and Yu.N. Ovchinnikov, Sov. Phys. JETP{\bf 20} (1965);
P. Fulde and R.A. Ferrel, Phys Rev. {\bf A135}, 550, 1964.}
\def\takada{S. Takada and T. Izuyama, Prog. theor. Phys. {\bf 41}, 635 (1969).}
\def\gaplessth{W.V. Liu, F. Wilczek, Phys. Rev. Lett{\bf{90}}, 047002 (2003);
S.T. Wu and S.K. Yip, Phys. Rev. {\bf 97},053603(2003).}
\def\gaplessexp{M.W. Zwierlein, A. Schirotzek, C.H. Schunck and W. Ketterle,
Science {\bf 311}, 492 (2006); G.B. Patridge, W. Li, Y. Kamer, R. LandLiao and
M.W. Zwierlein, A. Schirotzek, C.H. Schunck and W. Ketterle, cond-mat/0605258.}
\def\expbcsbec{K.M. O'hara {\em et al}, Science {\bf 298}, 2179 (2002);
C.A. Regal, M. Greiner and D.S. Jin, Phys. rev. Lett.{\bf 92}, 040403 (2004);
M. Bartenstein {\em et al}, Phys. Rev. Lett.{\bf 92}, 120401 (2004);
M.W. Zwierlein {\em et al}, Phys. Rev. Lett {\bf 92}, 120403 (2004);
J. Kinast {\em et al}, Phys. Rev. Lett. {\bf 92}, 150402 (2004);
T. Bourdel {\em at al}, Phys. Rev. Lett. {\bf 93}, 050401 (2004).}
\def\quarkmatter{For reviews see K. Rajagopal and F. Wilczek,
arXiv:hep-ph/0011333; D.K. Hong, Acta Phys. Polon. B32,1253 (2001);
M.G. Alford, Ann. Rev. Nucl. Part. Sci 51, 131 (2001); G. Nardulli,
Riv. Nuovo Cim. 25N3, 1 (2002); S. Reddy, Acta Phys Polon.B33, 4101(2002);
T. Schaefer arXiv:hep-ph/0304281; D.H. Rischke, Prog. Part. Nucl. Phys. 52,
197 (2004); H.C. Ren, arXiv:hep-ph/0404074; M. Huang, arXiv: hep-ph/0409167;
I. Shovkovy, arXiv:nucl-th/0410191.}
\def\htc{See e.g. in Q. Chen, J. Stajic, S. Tan and K. Levin,
Phys Rep {\bf 412}, 1 (2005) and references therein.}
\def\gas{G. Bertsch, in {\it Proceedings of the X conference
on Recent Progress in Many-Body theories,} edited by R.F.
Bishop {\em et al}.(World Scientific, Singapore, 2000).}
\def\rapid{B. Deb, A.Mishra, H. Mishra and P. Panigrahi,
Phys. Rev. A {\bf 70},011604(R), 2004.}
\def\crystal{M. Mannareli, K. Rajagopal and R. Shrma,{\PRD{73}{114012}{2006}};
M.G. Alford, K. Bowers and K. Rajagopal,{\PRD{63}{074016}{2001}};
R. Casalbuoni, R. Gatto, M. Mannarelli and G. Nardulli,{\PLB{511}{218}{2001}};
R. Casalbuoni, M. Cimanale, M. Mannarelli G. Nardulli,
M. Ruggieri and R. Gatto, {\PRD{70}{054004}{2004}}.}
\def\igor{Igor Shovkovy, Mei Huang, {\PLB{564}{205}{2003}}.}
\def\rischkeshov{M. Kitazawa, D.H. Rischke and I. Shovkovy,{\PLB{637}{367}{2006}}.}
\def\zhuang{L. He, M. Jin and P. Zhuang, Phys. Rev B{\bf 73},214527 (2006).}
\def\andreas{E. Gubankova, A. Schmitt, F. Wilczek, Phys. Rev. B{\bf 74}, 064505
(2006).}
\def\wilczek{E. Gubankova, W.V. Liu, F. Wilczek, Phys. Rev. Lett. {\bf 94}, 110402,
(2003).}
\def\hmparikh{H. Mishra and J.C. Parikh, {\NPA{679}{597}{2001}.}}
\def\amhma{Amruta Mishra and Hiranmaya Mishra,
{\PRD{69}{014014}{2004}.}}
\def\amhmb{A. Mishra and H. Mishra, {\PRD{71}{074023}{2005}.}}
\def\amhmc{A. Mishra and H. Mishra, {\PRD{74}{054024}{2006}.}}
\def\caldas{H. Caldas, arXiv:cond-mat/0605005}
\def\randeria{C.A.R. Sa de Melo, M. Randeria and J.R. Engelbrecht,
{\PRL{71}{3202}{1993}},{\PRB{55}{15153}{1997}}.}
\def\heiselberg{H. Heiselberg, Phys. Rev. A{\bf 63},043606 (2003).}
\def\tfd {H.~Umezawa, H.~Matsumoto and M.~Tachiki {\it Thermofield dynamics
and condensed states} (North Holland, Amsterdam, 1982) ;
P.A.~Henning, Phys.~Rep.253, 235 (1995).}
\def\amph4{Amruta Mishra and Hiranmaya Mishra,
{\JPG{23}{143}{1997}}.}
\def\rupak{P.F. Bedaque, H. Caldas and G. Rupak, 
{\PRL{91}{247002}{2003}}.}
\def\abrikosov{A.A. Abrikosov, L.P. Gorkov, Zh. Eskp. Teor.39, 1781,
1960.}
\def\krischprl{M.G. Alford, J. Berges and K. Rajagopal,
 {\PRL{84}{598}{2000}.}}
\def\marini{M. Marini, F. Pistolesi and G.C. Strinati, Eur. Phys. J.  
B{\bf 1}, 151(1998).}
\def\reddy{J. Carlson and S. Reddy,{\PRL{95}{060401}{2005}}.}
\def\son{Y. Nishida and D.T. Son,{\PRL{97}{050403}{2006}}.}
\def\ren{I. Giannakis, H. Ren,{\PLB{611}{137}{2005}};
{\em ibid}{\NPB{723}{255}{2005}}.}
\def\clog{A.M. Clogston, {\PRL{9}{266}{1962}}; B.S. Chandrasekhar,
Appl. Phys. Lett.{\bf 1},7, 1962.}
\def\amppn{ D. Silotri, D. Angom, A. Mishra and H. Mishra, in preparation.}
\def\loffitaly{R. Cassalbuoni and G. Nardulli, Rev. Mod. Phys. {\bf 76},
263,2004; M. Mannarelli, G. Nardulli, M. Ruggieri, arXiv:cond-mat/0604579}
\def\loffaust{X.J. Liu and H. Hu, Eur. Phys. Lett.{\bf 75},364 (2006);
H.Hu and X.J. Liu,{\PRA{73}{051603(R)}{2006}.}}
\def \hmspm {H. Mishra and S.P. Misra, {\PRD{48}{5376}{1993}.}}
\def\silotri{Pairing in spin polarised two species fermionic 
mixtures with mass asymmetry, Salman Silotri, Dillip Angom,
Hiranmaya Mishra  and Amruta Mishra, arXiv:0805.1784 (cond-mat) 
Eur. J. Phys.D49, 383-390 (2008).}
\def\koponen{M. Iskin, C.A.R. Sa de Melo, {\PRA{78}{013607}{2008}};
T.K. Koponen, T. Paananen,, J.-P. martikainen, P. Torma,
{\PRL{99}{120403}{2007}}.}

\end{document}